# Electrically-Detected ESR in Silicon Nanostructures Inserted in Microcavities

Nikolay Bagraev[a], Eduard Danilovskii[a], Wolfgang Gehlhoff[b], Dmitrii Gets[a], Leonid Klyachkin[a], Andrey Kudryavtsev[a], Roman Kuzmin[a], Anna Malyarenko[a], Vladimir Mashkov[c] and Vladimir Romanov[c]

[a]*Ioffe Physical Technical Institute, Polytekhnicheskaya 26, 194021 St. Petersburg, Russia*
[b]*Technische Universitaet Berlin, D-10623, Berlin, Germany*
[c]*Petersburg State Polytechnical University, Polytekhnicheskaya 29, 195251 St. Petersburg, Russia*

**Abstract.** We present the first findings of the new electrically-detected electron spin resonance technique (EDESR), which reveal the point defects in the ultra-narrow silicon quantum wells (Si-QW) confined by the superconductor δ-barriers. This technique allows the ESR identification without application of an external cavity, as well as a high frequency source and recorder, and with measuring the only response of the magnetoresistance, with internal GHz Josephson emission within frameworks of the normal-mode coupling (NMC) caused by the microcavities embedded in the Si-QW plane.

**Keywords:** Silicon microcavity, quantum well, ESR, point defects, negative-U center.
**PACS:** 72.25.Dc, 72.20.-i

## INTRODUCTION

Spin-dependent transport through semiconductor nanostructures in superconductor shells that are embedded in nano- and microcavities is of great interest to identify the magnetic resonance phenomena without using both the external cavity and the external hf sources and recorders [1, 2]. One of the best candidate on the role of such a 'sandwich' structure that is able to demonstrate the electrically-detected electron spin resonance (EDESR) by measuring the only magnetoresistance at high temperature appears to be the high mobility p-type silicon quantum well (Si-QW), 2nm, confined by the δ-barriers heavily doped with boron (Fig. 1). The findings of the electrical resistivity, thermo-emf and magnetic susceptibility measurements are actually evidence of the superconductor properties for these δ-barriers, 3 nm, $N(B)=5 \cdot 10^{21}$ cm$^{-3}$, which are revealed at critical sheet density of holes, $> 10^{11}$ cm$^{-2}$, in the p-type Si-QW on the n-type Si (100) surface [1, 2]. These δ-barriers have been shown to be type II high temperature superconductors (HTS) with $T_c=145$K and $H_{c2}=0.22$ T, thus providing the GHz and THz Josephson emission that can be enhanced by embedding the microcavity in the sandwich structure [2]. Here the S-Si-QW-S sandwich structures performed in the Hall geometry are used to register the ESR spectra of the point defects in p-type Si-QW using the EDESR technique above noticed.

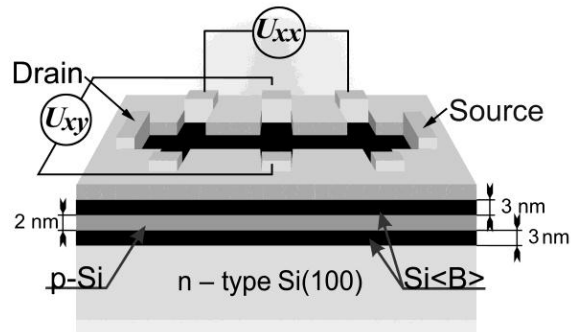

**FIGURE 1.** Schematic diagram of the device that demonstrates a perspective view of the p-type Si-QW confined by the δ-barriers heavily doped with boron on the n-type Si (100) surface. The depletion regions indicate the Hall geometry of leads.

## RESULTS AND DISCUSSION

The EDESR spectra of point defects revealed by measuring the only longitudinal or Hall magnetoresistance demonstrate the Rabi splitting that is attributed to the normal-mode coupling (NMC) with the single p-type Si-QW embedded in the silicon microcavity on the n-type Si (100) surface are shown in Figs. 2a, 2b and 3. Since the measurements of the magnetoresistance were performed without any light illumination and injection of carriers from the contacts, the EDESR effects appear to result from the spin-dependent scattering of spin-polarized holes on the paramagnetic centers in the edge channels of the S–Si-QW–S sandwich structures [2]. Therefore the value of the resonant magnetoresistance response appears to be interpreted here in terms of the interference transition in the diffusive transport of free holes respectively between the weak antilocalization regime ($\tau_S > \tau_\varphi > \tau_m$) in the region far from the ESR of a paramagnetic point defect located inside the edge channels and the weak localization regime ($\tau_\varphi > \tau_S > \tau_m$) in the nearest region of the ESR of that defect [2]. It should be noted that the important condition to register this resonant magnetoresistance response is to stabilize the drain-source current at the value of lower than 10 nA,

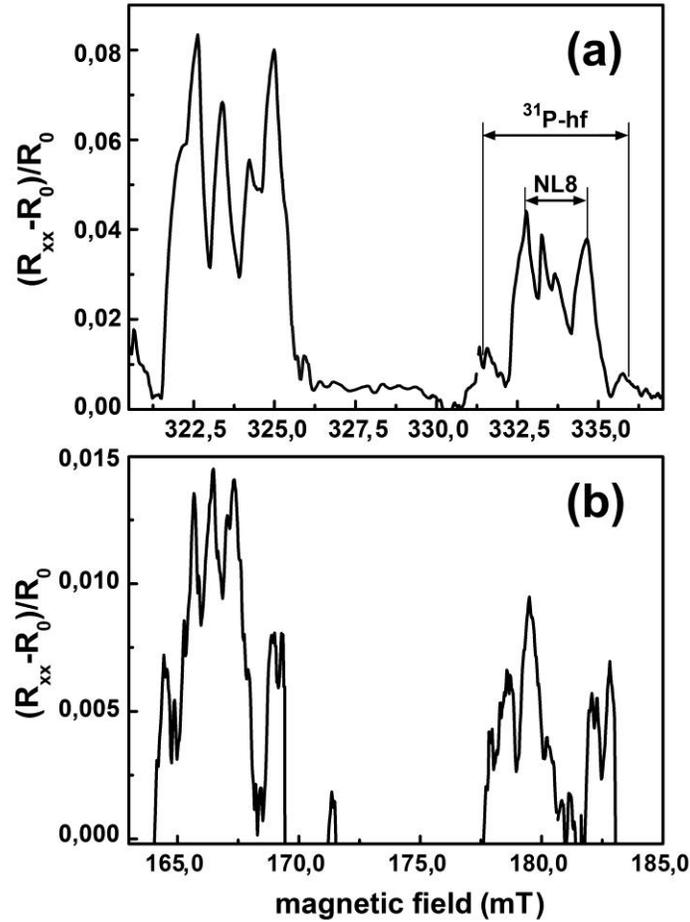

**FIGURE 2.** The EDESR spectra of point defects that reveal the Rabi splitting caused by the normal-mode coupling with the single p-type Si-QW embedded in the silicon microcavity on the n-type Si (100) surface.
(a) EDESR of phosphorus, NL8 and hydrogen-revealed centers in the Si-QW confined by the superconductor δ-barriers, which is observed by measuring a magnetoresistance without the external cavity as well as the hf source and recorder. T=77 K. B||<100> in the plane perpendicular to the {100} interface. ν=9.2 GHz. The 23 MHz splitting revealed by the central lines seems to be evidence of the hf hydrogen structure. (b) EDESR reply from the second harmonic.

which provides the spin interference regimes in the edge channels. Besides, the frequency selection that is controlled by measuring the Shapiro steps became it possible, because the special silicon microcavities have been embedded in the QW plane, with the sizes corresponding to the frequencies noticed above, L=λ/2n; here n is the refractive index,

L is the microcavity size and λ is the wavelength of the GHz- and THz-radiation [2]. Here the length of the δ-barrier area, L=4.79 mm, was picked up to satisfy to the selection of the frequency value of 9.2 GHz. Except the aforesaid it is necessary to pay attention that the EDESR spectra shown in Figs. 2a, 2b and 3 were observed at the temperature of 77K. This unusual result seems to be a consequence of sharp increase in spin-lattice relaxation time of holes in the low-dimensional silicon structures [1, 3].

The phosphorus ESR lines with the characteristic hf splitting of 4.1 mT are observed by measuring the only longitudinal magnetoresistance (Figs. 2a and 2b). The high sensitivity of the new EDESR technique is confirmed by the measurements of the NL8 spectrum that identifies residual oxygen thermodonors, TD+ - state, in the p-type Si-QW. This center of the orthorhombic symmetry has been also found by the ordinary EDESR method in the sandwich structure discussed here. The central lines in the EDESR spectrum are slightly different from the NL10 spectrum that is related to the neutral thermodonor containing a single hydrogen atom. Nevertheless, this EDESR spectrum appears to identify the hydrogen-related center in the p-type Si-QW, because its characteristic hf splitting, 23 MHz, corresponds to the hf hydrogen splitting. Different phases of the hf lines that result from the hydrogen-related center seem to be due to the high spin polarization of holes in 1-D channels [2, 3]. Besides, distinctive characteristic of the new type EDESR effect is additionally verified the observation of the same ESR spectrum under the conditions of the second harmonic (see Fig. 2b).

The high sensitivity of the EDESR technique allowed the studies in weak magnetic fields that are of importance for the measurements of the hf splitting for the centers inserted in the Si-QWs, which are characterized by the large g-values such as the $Fe^+$ center and the trigonal erbium-related center [2].

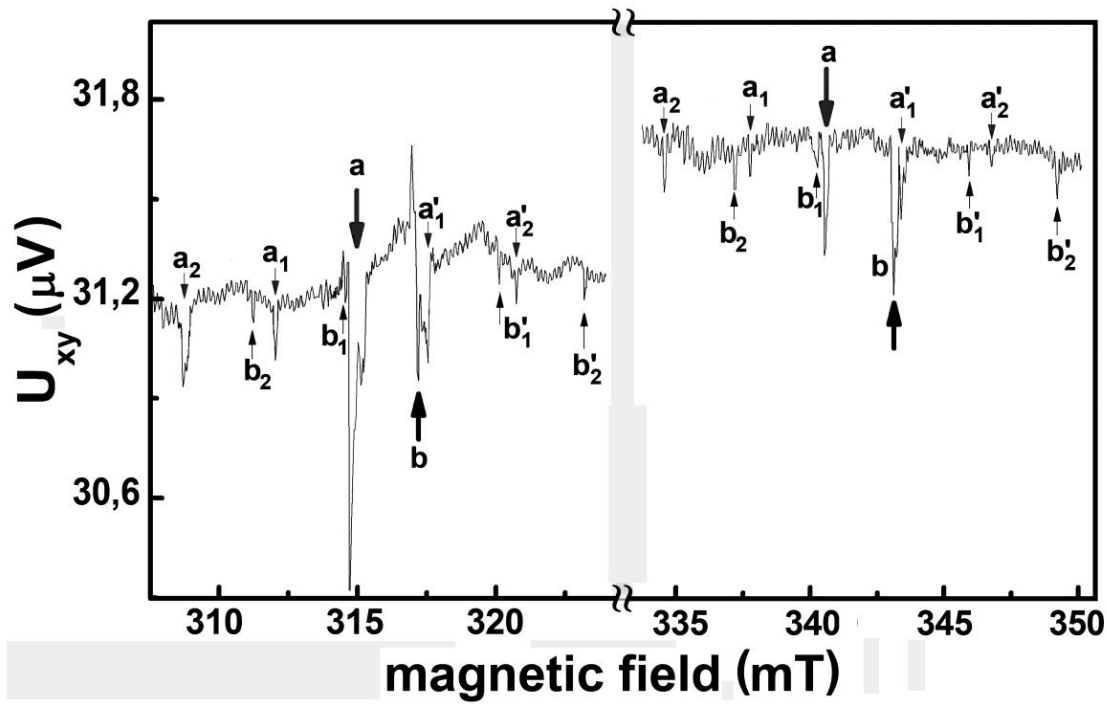

**FIGURE 3.** The EDESR spectra of isolated silicon vacancy that reveal the Rabi splitting caused by the normal-mode coupling with the single p-type Si-QW confined by the superconductor δ-barriers that is embedded in the silicon microcavity on the n-type Si (100) surface. This EDESR spectrum is observed by measuring a magnetoresistance without the external cavity as well as the hf source and recorder. T=77 K. B ∥ <100> in the plane perpendicular to the {100}-interface. ν =9.2 GHz. The pairs of the satellite lines appear to be related to the hyperfine interaction with the $^{29}Si$ nuclei.

The EDESR spectrum shown in Fig. 3 seems to be related to the $V_{Si}^+$ center, because following by George Watkins the identification of the ESR spectrum that results from the isolated silicon vacancy is caused by the presence of the $^{29}Si$ hyperfine satellites which are practically identical for a and b components and demonstrate the dangling bond character of the hyperfine interaction (see Fig. 4) [4]. Besides, these satellites exhibit the very close characteristics for both parts of the Rabi splitting (Fig. 3). However unlike classical behavior of the $V_{Si}^+$ center, the alignment is not revealed even at more high temperatures than 4.2K that seems to be due to the strain effects in the sandwich structure. The same reason seems to underlie the increased selectivity of the hyperfine interaction in the edge channels that appears to give rise to the additional number of the $^{29}$Si hyperfine satellites.

In spite of the negative-U properties of the $V_{Si}^+$ center, 2 $V_{Si}^+$ => $V_{Si}^{++}$ + $V_{Si}^-$, the EDESR spectrum shown in Fig. 3 was observed without preliminary illumination, hν < 0.35 eV, which is required for the registration of the classical ESR spectrum of the one-electron silicon vacancy state [4]. This result appears to be caused by the capture of holes from the edge channel in the course of the drain-source current and also consequence of existence of the dipole negative-U centers of boron. Finally, it is necessary to emphasize that the studies of this EDESR spectrum became it possible without any previous or subsequent e- irradiation, because the excess fluxes of the silicon vacancies were injected during all stages of the preparation of the silicon sandwich structures [1, 3].

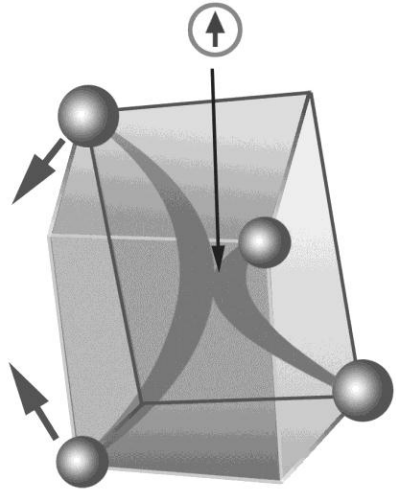

**FIGURE 4.** The model of the $V_{Si}^+$ center in silicon.

Thus, we present the findings of the new electrically- detected magnetic resonance technique (EDESR) which reveal single point defects in the ultra-narrow silicon quantum wells (Si-QW) confined by the superconductor δ-barriers. This technique allows the ESR identification without the application of the external cavity as well as the hf source and recorder by measuring the only magnetoresistance that reveals the Rabi splitting in the EDESR spectra within frameworks of the normal-mode coupling (NMC) caused by the microcavities embedded in the Si-QW plane.

## ACKNOWLEDGMENTS

The work was supported by the programme of fundamental studies of the Presidium of the Russian Academy of Sciences "Quantum Physics of Condensed Matter" (grant 9.12); programme of the Swiss National Science Foundation (grant IZ73Z0_127945/1); the Federal Targeted Programme on Research and Development in Priority Areas for the Russian Science and Technology Complex in 2007–2012 (contract no. 02.514.11.4074), the SEVENTH FRAMEWORK PROGRAMME Marie Curie Actions PIRSES-GA-2009-246784 project SPINMET.